\documentclass[journal]{vgtc}                     

\onlineid{1747}

\vgtccategory{Research}

\vgtcpapertype{evaluation}

\title{Embodied Natural Language Interaction (NLI): Speech Input Patterns in Immersive Analytics}

\author{%
  \authororcid{Hyemi\ Song}{0009-0004-5648-4478},
    \authororcid{Matthew\ Johnson}{0000-0002-7562-2926},
  \authororcid{Kirsten\ Whitley}{0000-0003-1356-326X}, 
  \authororcid{Eric\ Krokos}{0000-0003-1350-5297},
  \authororcid{Amitabh\ Varshney}{0000-0002-9873-2212}
}

\authorfooter{

  \item
  	Hyemi Song, Matthew Johnson, and Amitabh Varshney are with University of Maryland College Park.
  	E-mails: [hsong02 | mjohns28 | varshney]@cs.umd.edu.
  \item
  	Kirsten Whitley and Eric Krokos are with Department of Defense.
  	E-mails: [visual.tycho | ericpkrokos]@gmail.com
}

\abstract{%
    Embodiment shapes how users verbally express intent when interacting with data through speech interfaces in immersive analytics. Despite growing interest in \revision{Natural Language Interactions (NLIs)} for visual analytics in immersive environments, users’ speech patterns and their use of embodiment cues in speech remain underexplored. Understanding their interplay is crucial to bridging the gap between users’ intent and an immersive analytic system. To address this, we report the results from 15 participants in a user study conducted using the Wizard of Oz method. We performed axial coding on 1,280 speech acts derived from 734 utterances, examining how analysis tasks are carried out with embodiment and linguistic features. Next, we measured \revision{Speech Input Uncertainty} for each analysis task using the semantic entropy of utterances, estimating how uncertain users’ speech inputs appear to an analytic system. Through these analyses, we identified five speech input patterns, showing that users dynamically blend embodied and non-embodied speech acts depending on data analysis tasks, phases, and \revision{Embodiment Reliance} driven by the counts and types of embodiment cues in each utterance. We then examined how these patterns align with user reflections on factors that challenge speech interaction during the study. Finally, we propose design implications aligned with the five patterns.
}

\keywords{Embodiment, Natural Language Interaction (NLI), Immersive Analytics, Speech Patterns, Semantic Entropy, User Intent, Speech Acts}

\teaser{
  \centering
  \includegraphics[width=\linewidth, alt={A view of a city with buildings peeking out of the clouds.}]{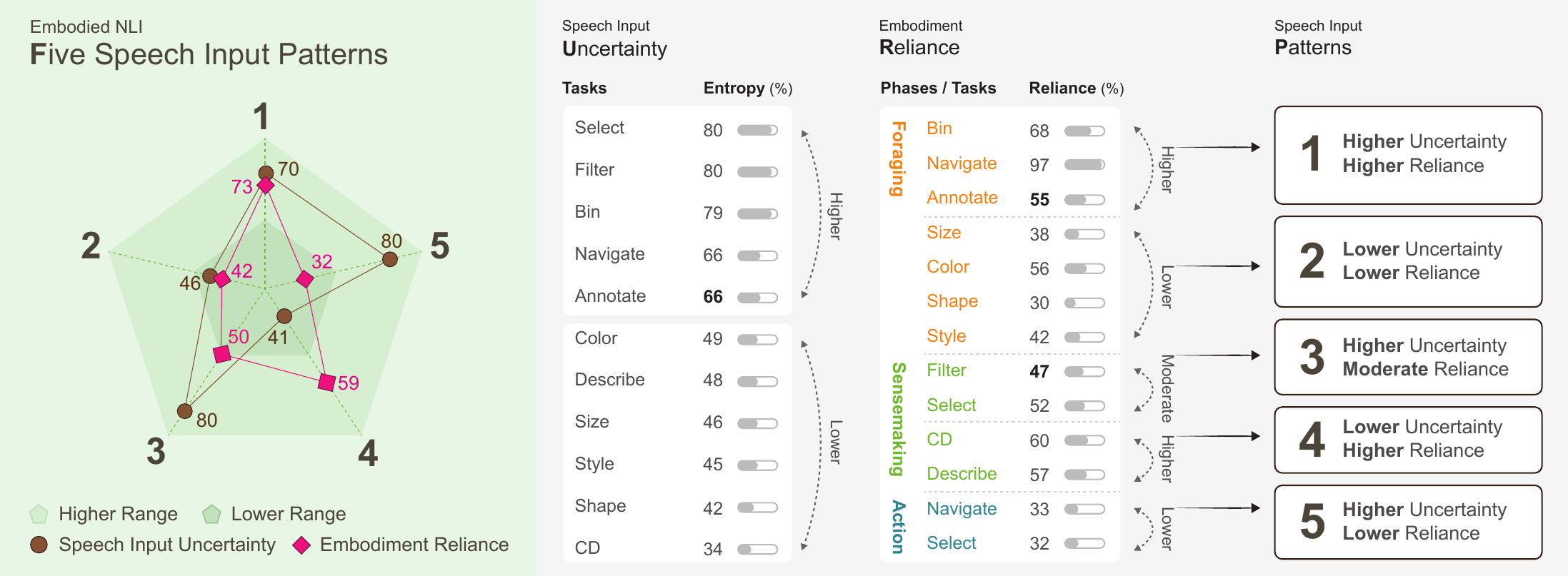}
  \caption{%
\textbf{[Left]} The chart summarizes the five speech input patterns. Each number represents the average percentage of Uncertainty or Embodiment Reliance among the tasks within a pattern (solely for illustrative purposes and not used in the analysis). \textbf{[Right]} This figure illustrates these five patterns, derived from Speech Input Uncertainty (measured as semantic entropy, scaled from 0 to 1, shown here as 0--100 for visual comparison) and Embodiment Reliance (percentage of embodied speech acts per task, 0--100\%). Uncertainty is categorized as \request {Higher} ($\geq$66) or \request{Lower} ($<$66). Reliance is grouped into \request{Higher} ($\geq$55), \request{Moderate} ($\geq$47 and $<$55), and \request{Lower} ($<$47). \revision{Boundary values (66, 55, 47) correspond to the maximum or minimum of each resulting cluster, which were determined by consecutively comparing the differences among all values to find the largest gaps}. CD denotes \revision{Characterize Distribution}.
  }
  \label{fig:teaser}
}

\graphicspath{{figs/}{figures/}{pictures/}{images/}{./}} 

\usepackage{tabu}                      
\usepackage{booktabs}                  
\usepackage{lipsum}                    
\usepackage{mwe}                       

\usepackage{mathptmx}                  

\graphicspath{{figs/}{figures/}{pictures/}{images/}{./}} 

\usepackage{colortbl}
\definecolor{lightgray}{RGB}{212,212,212}
\usepackage{xcolor}
\usepackage{colortbl}
\usepackage{tabularx}
\usepackage{placeins}
\usepackage{enumitem}
\usepackage[most]{tcolorbox}
\usepackage[skip=3pt]{caption}
\usepackage{afterpage}
\usepackage{makecell}
\usepackage[most]{tcolorbox}
\tcbset{
  highlight/.style={
    colback=gray!6,    
    colframe=gray!0,   
    boxrule=0pt,        
    arc=4pt,            
    left=3pt, right=3pt, top=0pt, bottom=0pt
  }
}

\newcommand{\todo}[1]{}
\newcommand{\addcite}[1]{}
\newcommand{\request}[1]
{\textcolor{black}{#1}}
\newcommand{\revision}[1]{\textcolor{black}{#1}}

\newcommand{\eg}{{e.g.,}\xspace}
\newcommand{\ie}{{i.e.,}\xspace}
\newcommand{\etal}{{et al.}\xspace}
\newcommand{\bpstart}[1]{\noindent{\textbf{#1.}}}

\begin{document}



\maketitle


\section{introduction}
Recent advances in immersive analytics and generative AI have fueled a growing interest in natural interactions between humans and data analytic systems through Natural Language Interactions (NLIs) \cite{hoque_chart_2022, kim_explainable_2025}. Previously, NLIs for traditional visualization mainly focused on the linguistic perspective, particularly textual interactions on the desktop, and did not account for embodiment \cite{metoyer_understanding_2012, srinivasan_collecting_2021, tory_what_2019, gao_datatone_2015, setlur_eviza_2016, cao_dataparticles_2023, tang_sevi_2022, wu_nl2viz_2022, wang_towards_2023}. More recently, generative AI-infused visual analytics and authoring tools have gained attention. However, they tend to focus on model capabilities or how NLIs are converted into system responses \cite{tian_chartgpt_2025, dibia_lida_2023, wang_dracogpt_2025, cui_promises_2025}, often assuming that users naturally use speech without a deeper investigation into their speech behaviors.

In immersive analytics \cite{ens_grand_2021, zhang_survey_2023, liu_interactive_2022, marriott_immersive_2018, fonnet_survey_2021}, speech is inherently more natural than textual input, and embodiment is often used to convey intent while physically exploring elements. In these processes, speech and embodiment interact in sophisticated ways. For example, when users analyze data using speech in immersive environments, they frequently incorporate embodiment cues into their utterances (examples: \hyperref[tbl:Hierarchy]{Table 1}; definitions: \hyperref[chap:terms]{Section 2.4}). We introduce this phenomenon and  refer to it as \textit{Embodied NLI}. Although previous work has recognized the complementary relationship between speech and gesture \cite {bolt_put-that-there_1980, mcneill_speech_1998, goodwin_action_2000, batch_wizualization_2024}, no research has thoroughly examined users' speech behaviors and how embodiment cues function within them in immersive analytics.

In addition, understanding the intricate interactions between speech and embodiment factors (\ie Embodiment Reliance, Cue Type and Cue Count) is important for designing NLI-based immersive analytics tools. Such understanding enables the system to accurately and proactively interpret users' intent and related actions by using NLIs, optimizing the capabilities of generative AI assistants. As a result, this study aims to effectively bridge the gap between users' intent and immersive analytic systems' outcomes by addressing two primary research questions:

\begin{itemize}[noitemsep,topsep=1pt,parsep=1pt,partopsep=0pt]
\item \textbf{RQ1}: What types of utterances do users produce when interacting with immersive visualizations? How are these utterances related to their data analysis tasks, intents, and embodiment factors?
\item \textbf{RQ2}: Which interaction and interface factors affect users' challenges in speech-based interaction with immersive visualizations?
\end{itemize}


\vspace{-2mm}
\captionsetup{belowskip=-10pt}
\begin{table}[h]
\request{
    \centering
    \begin{tabular}{p{3.6cm} p{1.9cm} p{2.2cm}}
    \toprule
    Utterance & Speech Act & Embodiment Cue\\
     \midrule
    Basically, the client I’m helping has a limited budget, so I’d like to select these data points on the left side of me and color them red.&
    1) Select these data points.\newline 
    2) Color them red.&
    1) These\newline 
    2) Them\newline
    3) On the left side\\
     \bottomrule
    \end{tabular}
    \caption{Examples: one utterance, two speech acts, three cues.}
    \label{tbl:Hierarchy}
} 
\end{table}

We conducted a Wizard of Oz based user study in a Virtual Reality (VR) environment to collect utterances and users' self-reported challenges encountered during the study.
 We then performed qualitative analysis (\ie open and axial coding) and measured \textit{Speech Input Uncertainty} from the collected data. Based on these analyses, we identified five speech input patterns by examining relationships between uncertainty levels and \textit{Embodiment Reliance}, defined as users’ reliance on embodiment cues during data-analysis tasks. The identified patterns illustrate how users dynamically blend embodied and non-embodied speech acts depending on the analysis task (\eg Filter), analysis phase (\ie Foraging, Sensemaking, Action), and embodiment factors.
 To validate these patterns, we grounded our analysis in user reflections by referencing their comments. These discussions became the foundation for our proposed design implications.

The contributions of this paper are: \begin{itemize}[noitemsep,topsep=1pt,parsep=1pt,partopsep=0pt] 
\item C1: Introducing terminology, definitions, and research methods for Embodied NLI, including embodiment cues, embodied and non-embodied speech acts, and Speech Input Uncertainty. 
\item C2: Proposing a speech-input analysis framework, outlining how analysis tasks are accomplished by performing strategies and verbalizing utterances, thus transferring Speech Input Uncertainty to immersive analytic systems. 
\item C3: Identifying five speech input patterns across tasks and phases based on Speech Input Uncertainty and Embodiment Reliance. 
\item C4: Formulating design implications for speech-based immersive analytic tools. 
\end{itemize}

\section{Background and Related Work}
\subsection{Natural Language Interface and Interaction (NLI) for Data Visualization} 

Efforts to bridge the gap between users' intent and computer systems through NLIs have long been ongoing since early research in Human-Computer Interaction (HCI) \cite{bush_as_1945, licklider_man-computer_1960, simon_science_design_1996}.


\request{
However, NLIs have generally not been considered a primary input modality in HCI. Hutchins \etal \cite{hutchins_direct_1985} showed that the interface as conversation introduces both semantic distance (between intentions and interface language) and articulatory distance (between input actions and that language), while direct manipulation minimizes these distances by reducing the gulfs of execution and evaluation. These distances increase users’ cognitive load and raise the risk of misinterpretation or erroneous commands. Therefore, it is not surprising that existing studies in visualization have also treated voice primarily as a supplemental input channel. While some work still tackles NLI-based visualization challenges \cite{besancon_state_2021, srinivasan_nli-vis_2017}, recent advances in Large Language Models (LLMs) are rapidly closing the gap.
}

\request{
Together with efforts to apply NLIs to data visualization, high demand persists for visualization (vis)-centric approaches, as general HCI and XR research often overlook vis-specific interaction tasks and chart elements. For example, Besançon \etal \cite{besancon_state_2021} surveyed interaction paradigms for 3D visualization, and Dimara \etal \cite{dimara_what_2020} examined definitions and approaches to interaction in data visualization. Regarding NLIs, Hoque \etal \cite{hoque_chart_2022} recently identified a shortage of studies that treat natural language as both an input and an output in data visualization. With the advent of LLMs, visualization researchers have now distinguished between LLM-based methods for generating charts and generic artifacts. This paper aligns with these directions and examines vis-specific speech-input patterns along with data analysis tasks.
}

\request{Even before LLMs became popular, researchers studied NLIs for traditional data visualization, focusing on linguistic aspects, particularly for textual queries and responses.} Systems such as Eviza \cite{setlur_eviza_2016} and DataTone \cite{gao_datatone_2015} explore the ambiguity of user textual queries by highlighting ways to clarify user intents. Additionally, a toolkit \cite{narechania_nl4dv_2021} and NLI-based chart generation studies facilitate the conversion of text inputs into visualizations \cite{tian_chartgpt_2025, dibia_lida_2023, wang_dracogpt_2025, cui_promises_2025, feng_xnli_2024, tang_sevi_2022, wu_nl2viz_2022}. These linguistically-focused approaches are effective in traditional visualization contexts. However, immersive visualizations demand a fundamentally different approach due to the inherent characteristics of an immersive environment, in which speech and embodiment become natural interaction modes.

Beyond the desktop context, early influential works such as Put That There \cite{bolt_put-that-there_1980} \revision{demonstrated} how speech and gesture complement each other. Similarly, visualization studies suggest that multimodal inputs (\eg speech and touch, speech and gestures) effectively help users express analytic intent \cite{srinivasan_orko_2018, srinivasan_inchorus_2020}; however, previous work does not deeply focus on speech. For example, systems such as ImAxes \cite{cordeil_imaxes_2017} and Data Visceralization \cite{lee_data_2021} \revision{emphasized embodiment but did} not explicitly investigate speech. Meanwhile, the existing design frameworks provide theoretical and empirical guidelines for embodiment \cite{marriott_just_2018, saffo_unraveling_2024}; there has been little discussion about speech.

\request{Although speech is studied more in Extended Reality (XR) than in immersive analytics, studies still mainly evaluate interaction techniques and system performance \cite{in_evaluating_2024, piumsomboon_grasp-shell_2014}, rarely treating speech as the primary input and seldom examining users’ speech behavior in depth.
}

Nevertheless, to our knowledge, no prior literature has conducted an in-depth analysis of users' speech behaviors or examined how embodiment cues are expressed in speech within immersive analytics. This paper addresses this gap by identifying five speech input patterns and related insights to guide future intelligent visualization systems.

\subsection{Speech, Embodiment, and Intent}
\bpstart{Speech Act}
An utterance is a widely used linguistic unit encompassing spoken and written language. With this terminology, we introduce the notion of \textit {Speech Act} to capture user intent from spoken utterances more closely. This term was introduced by English philosopher John Austin via his theory of speech acts, emphasizing that it represents an intended action rather than a grammatical component \cite{austin_how_1975}. An utterance is a higher-level unit that may include multiple speech acts, ranging from individual words to phrases and complete sentences. Extending this concept, we adopt the \textit {Speech Act} as the unit of analysis in this paper, defining it as the smallest unit of speech that conveys the speaker's intended action to modify a chart or data element. For example, given an utterance, \revision{
\textit{"I’d like to select these
data points on the left side
of me and color them red."} a speech act is \textit{"Select these data points."}
}

\bpstart{Speech-Gesture Interconnection and Embodiment} As in the definition of \textit {Speech Act}, speech can express intent for action. Similarly, the interconnection between gesture and speech has been extensively studied. McNeill \cite{mcneill_speech_1998} argues that these modalities originate from a common psychological structure, which allows them to mutually express the speaker's intent. McNeill highlights the cognitive connection between \revision{these modalities}. Meanwhile, Goodwin \cite{goodwin_action_2000} offers a different perspective, revealing how these modalities are dynamically co-constructed within embodied interactions shaped by environmental context.

Based on these perspectives, a deeper examination of embodiment is necessary to understand how speech and gesture jointly shape verbal communication and how embodiment factors influence this process. Kilteni et al. \cite {kilteni_sense_2012} define the Sense of Embodiment (SoE) in XR, emphasizing three key factors that affect immersive experiences: self-location (spatial awareness from the user's perspective), agency (control over speech and gesture), and ownership (the self-attribution of one's body). These factors provide a useful lens for identifying embodied behaviors associated with speech, such as the types of speech acts (\ie Embodied and Non-Embodied Speech Acts). We use these approaches when designing the axial coding scheme described in \hyperref[chap:uncertainty]{Section 4.4}.

\subsection{From Uncertainty to Speech Input Uncertainty} \label{chap:bg_uncertainty}
Beyond understanding users' speech and embodied interactions, our central question is how an immersive analytic system interprets users' speech inputs. These inputs are closely tied to the system's internal processes and resulting outputs. For example, uncertain inputs expand the range of probable outcomes by increasing the information entropy that the system must \revision{process}. This cause-and-effect relationship motivated us to propose identifying speech input patterns based on approaches for quantifying \revision{Speech Input Uncertainty}, thereby examining how intelligent immersive analytic systems can interpret user speech input.

\request{\textbf{Uncertainty} refers to \textit{the state of lacking definite knowledge or clarity}, which has been studied across various domains \cite{bhatt_uncertainty_2021, baan_uncertainty_2023}. In machine learning, researchers define it as \textit{the state of our lack of knowledge about model outcomes}. To reason about unknown model behavior, researchers have long employed \textbf{uncertainty quantification}, enabling the assessment of the reliability and trustworthiness of model outputs \cite{duan_shifting_2023, lin_generating_2024, huang_look_2025, bhatt_uncertainty_2021}. Depending on the task type, the quantification methods are applied differently (\eg confidence intervals for regression tasks and entropy for classification tasks). With this trend, uncertainty visualization has also gained attention as a means to efficiently interpret statistical outputs in data visualization \cite{hullman_why_2020, wilke_fundamentals_2019, kay_when_2016}.}

\request{As LLMs have recently become popular, uncertainty quantification in natural-language generation (NLG) has garnered growing interest \cite{baan_uncertainty_2023, kuhn_semantic_2023}. In this context, researchers define uncertainty in NLG as \textit{the state of our lack of knowledge about the probability that an LLM will generate a specific utterance.} When measuring uncertainty, they calculate entropy—because sentence-based predictions are treated as classification tasks—by considering the similarity between utterances.} 

\request{Previous work treats uncertainty from the human interpreter's perspective. Now we flip the lens: the system itself becomes the interpreter of human speech. We refer to this system-centered perspective as \textbf{Speech Input Uncertainty}, which is \textit{the state of a system’s lack of knowledge in mapping a given utterance to a specific task (\eg Filter).} We quantify this lack of knowledge as a probability.}

\request{Despite growing interest in NLG, the existing literature primarily estimates LLM uncertainty using syntactic or lexical similarity, overlooking semantic similarity, which can be conveyed by phrases that are very different yet convey the same meaning—for example, \textit{``France's capital is Paris''} or \textit{``Paris is France's capital.''} Recent work, such as that by Kuhn et al. \cite{kuhn_semantic_2023}, addresses this gap by quantifying semantic entropy to measure \textbf{semantic uncertainty} in LLMs through three steps: 1) generating sentence samples from an LLM, 2) clustering semantically equivalent samples, and 3) computing semantic entropy to measure uncertainty over those clusters. We leveraged this method to analyze human speech in immersive analytics: by treating humans as a "human language model", we 1) collected their utterances, 2) clustered them by meaning, and 3) measured semantic entropy. The detailed methodology is presented in \hyperref[chap:uncertainty]{Section 4.4.2}.}

\subsection{Terminology and Definitions} \label{chap:terms}
We introduce additional key terms and definitions. 
\begin{itemize} [noitemsep,topsep=1pt,parsep=1pt,partopsep=0pt]
   \item \textbf{Embodiment Cue} is a word or phrase that conveys pointing or spatial context, such as \textit{these} and \textit{on the left} (\eg Select these data points on the left side of me). It is a distinct linguistic element within a speech act but does not independently establish full contextual meaning. \hyperref[tbl:Hierarchy]{Table 1} shows examples.
   \request{
   \item \textbf{Embodied Speech Act} is a speech act that contains at least one embodiment cue (\eg Select \underline{these} data points: the cue type of "these" is Deictic-Single, see \hyperref[tbl:dimension]{Table 2}). In contrast, \textbf{Non-Embodied Speech Act} does not include such cues (\eg I would choose House ID 67).}
   \item \textbf{Analysis Phase or Phase} stems from the \revision{{sensemaking loops}} by Pirolli and Card \cite{pirolli_card_sensemaking_2005}. We introduce an additional step, the \textit{Action} phase, which extends the original \revision{sensemaking loops} by explicitly distinguishing between an internal selection decision and an external execution. While selections in the \textit{Foraging} and \textit{Sensemaking} phases represent internal cognitive decisions, the \textit{Action} phase captures how users communicate these internally formulated decisions as explicit, detectable signals to the system.
    \item \textbf{Analysis Task or Task} refers to data analysis tasks for immersive analytics. Details are discussed in \hyperref[chap:analysismethods]{Section 4.4}.
    \item \textbf{Strategy} informs an actionable approach to accomplishing a given task through verbalizing an utterance. Each task entails multiple strategies. Details are explained in \hyperref[chap:uncertainty]{Section 4.4.2}.
    \item \textbf{Embodiment Reliance} refers to the proportion of embodied speech acts among all speech acts used when performing an analysis task (\eg Filter).
    \item \textbf{Embodiment Cue: Count} indicates the number of embodiment cues expressed in a speech act. We categorize them into three levels: multiple cues, single cues, and no embodiment cues.
    \item \textbf{Embodiment Cue: Type} refers to the classification of embodiment cues into one of seven categories: \textit{Deictic-Single} (\eg these), \textit{Deictic-Boundary} (\eg from here to this line), \textit{Deictic-Time} (\eg undo all filters), \textit{Spatial-Direction} (\eg on the left side of them), \textit{Spatial-Distance} (\eg far from me), \textit{Spatial-Body} (\eg move closer to the chart), and \textit{Spatial-Area} (\eg fall into this category).
\end{itemize}



\vspace{-1mm}
\section{Pilot Study and Observations} \label{chap:pilot}
After developing a VR application using Unity, we conducted a pilot study with four participants. The study consisted of two sessions: a 20-minute \revision{structured task session,} and a 30-minute open-ended analysis. In this scenario, participants role-played as realtors tasked with finding the best house for their client by interacting with a 3D scatterplot while wearing a Head-Mounted Display (Meta Quest Pro). We identified several key findings from the study:
\begin{itemize}[noitemsep,topsep=1pt,parsep=1pt,partopsep=0pt]
\item F1: Participants actively and passively used gestures to express their
intent while talking to the system.
\item F2: Participants primarily used a few tasks, such as filtering during exploratory analysis.
\item F3: Participants with prior visualization expertise interacted with the system more efficiently.
\item F4: Participants familiar with NLP expected the system to handle
more complex reasoning.
\item F5: Some participants, recalling communication challenges with AI, tried simpler expressions, raised their voices, and made dramatic gestures.
\item F6: All participants misunderstood the study as evaluating an AI system, so they focused on very low-level tasks.
\item F7: As the study progressed, participants expressed their intent using more consistent language.
\end{itemize}

\noindent Based on these observations, we formulated two assumptions:
\begin{itemize}[noitemsep,topsep=1pt,parsep=1pt,partopsep=0pt]
\item A1: People may alter how they speak when combining speech and gestures for data analysis in immersive analytics (F1, F2).
\item A2: People's interactions with the visualization might evolve throughout the analysis session, influenced by learning, practice, and task complexity (F1, F2, F7).
\end{itemize}
These findings relate to our two research questions. Two assumptions are further discussed in \hyperref[chap:initial_analysis]{Section 5.1}.


\vspace{-1mm}
\section{methods} \label{chap:method}
\subsection{Study Design}
Informed by the pilot study findings in \hyperref[chap:pilot]{Section 3}, we adjusted the main study procedure. The study began after obtaining IRB approval. Each participant’s session included an introduction, a consent form, three steps (Passive Analysis, Tutorial, Active Analysis), and an exit interview. Each study lasted no more than 70 minutes. The first and second authors participated: the second author served as the interviewer, actively communicating with participants, and the first acted as the Wizard, controlling the wizard system and managing recordings.

During the passive analysis, \revision{participants were shown two 3D scatterplots—each depicting a real-estate dataset—with one or two visual modifications (e.g., color).} Participants were asked to brainstorm and verbalize how to make these changes by requesting the refinements from an editor (an intelligent system or a human editor). During the pilot study, we found that participants relied heavily on a few tasks (\eg Filter), resulting in an unbalanced collection of utterances. To resolve the problem, we asked participants to focus on less frequently used tasks (\eg Encode). We also emphasized that the goal was to capture participants' natural expressions, not to evaluate an AI system.

In the tutorial session, participants read five example utterances (\eg Show me the top 10\% affordable houses), with the Wizard updating visualizations accordingly. The final utterance was intentionally complex and not executable, prompting a warning message to highlight system limitations and potential resolutions.

After learning how to interact with the editor, participants engaged in a role-play scenario as realtors tasked with finding the best house for their clients. The interviewer provided \revision{instructions} and the client's profile and preferences, which they could reference via a panel in the headset. Eventually, the active \revision{analysis} was an open-ended exploration. \revision{After participants confirmed their final choices, the interviewer conducted a verbal interview and then administered a written questionnaire.}


\subsection{Participants}
We recruited participants by advertising the study both online (e.g., Slack, email) and offline (via flyers). The recruitment criteria were 1) Be at least 18 years old, 2) Be able to arrive at the study place to participate in this study, 3) Have experience using charts in data analysis, 4) Have normal (or corrected) vision, and 5) Be fluent in English. Most participants had at least basic knowledge of data analysis and minimal or no experience in XR. The 15 participants in our study ranged in age from 18 to 35, and all participants had at least a high school diploma (some held a bachelor’s or master’s degree). 

\subsection{Application, Recording, and Data Collection}
\bpstart{Application}
We adopted a Wizard of Oz approach for the entire study. To implement this, we developed an application in Unity (v2021.3.4f1) using IATK \cite{cordeil_iatk_2019} on a Meta Quest Pro. We used \revision{a wired Quest Link} to establish a stable connection between the central system and the Quest Pro. The application has two views: a participant view and the wizard's view. The former presents a 3D scatterplot and UI panels in an egocentric perspective, and the latter displays the chart from a bird's-eye view in a separate window. On the wizard side, we also extended the Unity Inspector window with additional UI components. To test various interactions on the Scatterplot, we extended IATK by adding interaction and data transformation functions. Finally, we integrated these features into the extended Inspector UI. These windows help the wizard (i.e., the interviewer) respond quickly to participants' requests.

\noindent\request{
\bpstart{Scatterplot} \label{chap:scatterplot}
Following Sadana and Stasko \cite{sadana_designing_2014}, who intentionally narrowed their focus to scatterplots to enable in-depth exploration, this study focused exclusively on scatterplots. We selected this chart type because it is the most widely used in information visualization systems and sufficiently supports the investigation of various interactions \cite{sarikaya_scatterplots_2018}.}
\bpstart{Recording}
We recorded participants' behavior and wizard's computer screens (including the Unity Inspector) using cameras (one Webcam, one Quest Pro), two mics, three computers, and four monitors.

\bpstart{Data Collection and Digitization}
We collected all utterances from videos, audio, and transcripts, as well as self-reported responses from exit interviews. These utterances were then categorized by type (\ie request, interpretation, brainstorming, intent, strategy, and final choice) and by three analysis phases: Foraging (request, interpretation), Sensemaking (brainstorming, strategy, intent), and Action (final choice). In total, 734 utterances (1,280 manually extracted speech acts) were gathered from 15 participants, along with their self-reported responses.


\subsection{Analysis Methods} \label{chap:analysismethods}
\request{
We employed two methods: 1) qualitative analysis, which examines the utterance types users employ to express intent in speech by incorporating embodiment cues and identifies the challenges and strengths encountered in the study; and 2) \revision{Speech Input Uncertainty} measurement, which quantifies semantic uncertainty across collected utterances.
}
\vspace{-3mm}
\subsubsection{Qualitative Analysis}
\request{
\bpstart{Hybrid Axial Coding: Utterance-Type Identification} 
We coded speech acts into three analysis phases, eleven analysis tasks, and seven embodiment cue types. The analysis phases are based on Pirolli and Card \cite{pirolli_card_sensemaking_2005}. The analysis tasks are drawn from Marriott et al.'s \cite{marriott_just_2018} "How" category, as our goal is to examine how users verbalize their intent. From the original "How" category, which comprises Encode, Manipulate, Facet \& Position, Reduce, Collaborate, Render, and Model, we excluded Facet \& Position, Collaborate, and Render because we focus on single users' interaction behaviors. Additionally, we added Explain (\ie elaborating data) and Styling (\ie visual elements without data binding such as panel and line color) to capture users' communicative behaviors. We then defined subcategories (e.g., Explain: Annotate, Describe) referring to other taxonomies \cite{amar_low-level_2005, brehmer_multi-level_2013}.}

\request{
Marriott et al.'s "Why" category maps to user intent, which we did not code directly. Rather, we treated intent as a higher-level goal (e.g., Explore) realized through an analysis task (e.g., Filter). After initial analysis, we excluded subcategories with fewer than 30 speech acts to ensure statistical reliability in measuring \revision{Speech Input Uncertainty}. Finally, we applied eleven analysis tasks, including Encode (Size, Shape, Color), Reduce (\revision{Bin}), Model (Characterize Distribution), Manipulate (Filter, Select, Navigate), Explain (Annotate, Describe), and Style. 
}

\request{
For embodiment cues, we drew on Kilteni \etal \cite {kilteni_sense_2012}, which defines three factors influencing the sense of embodiment: Ownership, Self-Location, and Agency. The latter two became the basis for the cues' categories as shown in \hyperref[tbl:dimension]{Table 2}. Using these categories, we tagged each speech act with one task (\eg \revision{Bin}), single or multiple cues (\eg Above, Those), and one phase (\eg Foraging). For example:
}
\vspace{-1mm}
\noindent\begin{tcolorbox}[highlight]
  \textit{``\underline{Above} the mean price, you paint \underline{those} data points as blue\ldots''} \\
  (Task: \revision{Bin}, Phase: Foraging)
\end{tcolorbox}
\vspace{-1mm}

\vspace{-2mm}
\begin{table}[h]
\request{
  \centering
  \begin{tabular}{p{2cm} p{2cm} p{3.4cm}}
    \toprule
    Phase (3) & Task (11) & Embodiment Cue: Type (7)\\
    \midrule
    Foraging\newline Sensemaking\newline Action
      &
    Encode (3)\newline
    Reduce (1)\newline
    Model (1)\newline
    Manipulate (3)\newline
    Explain (2)\newline
    Styling (1)
      &
    {%
      \raggedright
      Deictic:\\
      \hspace{0.5em}Single, Boundary, Time\\[0.5ex]
      Spatial:\\
      \hspace{0.5em}Direction, Distance,\\
      \hspace{0.5em}Body, Area
    }\\  
    \bottomrule
  \end{tabular}
}  
  \caption{Coding categories, subcategory counts in parentheses.}
  \label{tbl:dimension}
\end{table}


\noindent\request{
\bpstart{Thematic Analysis: Challenges and Strengths}
We analyzed the written and oral interviews concerning the challenges and strengths of the study’s approaches, following a thematic analysis. All responses were first coded into multiple themes and then iteratively consolidated, ultimately yielding categories that address RQ2 in \hyperref[chap:RQ2]{Section 6.3}.
}%

\bpstart{Analysis Process}
The first author conducted \revision{open and axial coding}, guided by prior literature \cite{saldana_coding_2021, mcdonald_reliability_2019}, and refined the codebook, code \revision{scheme}, and results through iterative discussions with co-authors. To further support reliability, we also used LLMs (Claude v3.7 Sonnet, GPT-4) as auxiliary validators, providing prompt-based instructions via chat. These models were not involved in the coding itself but were used only to determine the necessity of more iterations and discussions with co-authors. Discrepancies between \revision{the authors} and model outputs were reviewed in-depth by the first author. Critical issues were further discussed with co-authors. These processes were iteratively applied until consistency across the results was clearly established.


\subsubsection{Speech Input Uncertainty Measurement} \label{chap:uncertainty}


\request{
\bpstart{Data Preparation}
Before measuring uncertainty, we grouped speech acts by task (e.g., Filter: 207 speech acts). To enhance the contextual understanding of a NLI model for measuring entropy, we paired each speech act with its corresponding utterance, forming tuples. Using the tuples, we measured Speech Input Uncertainty per task, quantified as semantic entropy. Our approach follows the method proposed by Lorenz Kuhn et al. \cite{kuhn_semantic_2023}, which involves three steps \hyperref[chap:bg_uncertainty]{(see Section 2.3)}.
}

\noindent\request{
\bpstart{Entropy Measurement}
The first step was to collect utterances and prepare tuples. We then clustered the tuples for each task by computing entailment scores between pairs of tuples using a Natural Language Inference (NLI) model, specifically RoBERTa-large-MNLI \cite{facebookai_roberta-large-mnli_nodate}, noted for its high cross-domain accuracy. After forming clusters, we computed the probability of each cluster. In the original paper \cite{kuhn_semantic_2023}, cluster probability is defined as the likelihood that an LLM will confidently generate a given sentence. In our paper, we measured the cohesiveness of human speech expressions by considering a random group of users as a human language model and estimating the probability that multiple users would ultimately verbalize similar expressions within a cluster to perform a task. Entropy depends on two factors. First, the number of clusters (e.g., ten clusters) per task (e.g., Filter); more clusters mean higher entropy, indicating greater uncertainty in verbal expressions (i.e., tuples) used to perform a task. Second, the cohesion of the expressions within a cluster (e.g., a cluster within filter); lower cohesion results in higher entropy. This cohesion is derived using the cosine distance between individual expressions and their cluster centroid in the embedding space. The relative distances were used to compute each cluster’s probability. After computing the probabilities for all clusters within a task, we calculated the entropy by passing the list of cluster probabilities to the entropy function in SciPy \cite{scipy_community_scipy_nodate}. Finally, the results were normalized to scores between 0 and 1 for each task. 
}

\subsubsection{Speech Input Analysis Framework}
After measuring entropy, we qualitatively analyzed clusters' speech acts to extract representative themes defined as user strategies to perform an analysis task. Based on this analysis, we developed an analytical framework that systematically connects a task (\eg Filter), a strategy (\eg Elimination), an utterance (including speech acts and embodiment cues), and quantified \revision{Speech Input Uncertainty}, illustrating how the computed uncertainty is transmitted into immersive analytic systems. The identified top-N strategies are summarized in \hyperref[tbl:Semantic Clusters]{Table 3}. We apply this framework to analyze the speech input patterns in \hyperref[chap:results]{Section 6}.

\vspace{-2mm}
\begin{figure}[!ht]
  \centering
  \begin{subfigure}[b]{1\columnwidth}  
  	\centering
  	\includegraphics[width=\textwidth, alt={Big letter A on a gray background.}]{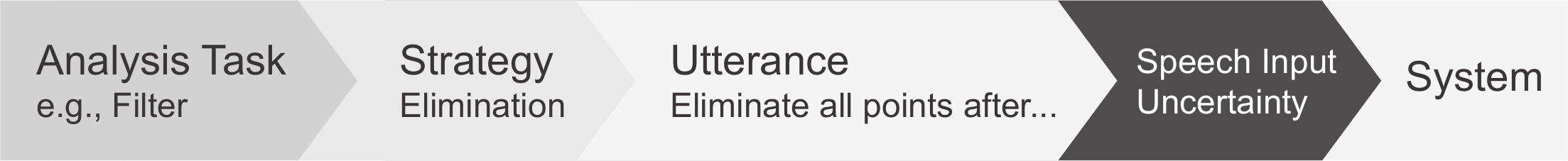}
  	\label{fig:framework}
  \end{subfigure} 
\vspace{-6mm}
  \caption{Speech Input Analysis Framework.}
\end{figure}

\afterpage{%
\begin{table*}[h]
    \centering
    \small
    \begin{tabular}{p{4cm} p{13cm}}
    \toprule
     Analysis Tasks & Top-N Strategies\\
     \midrule
      Size  & Conditional Scaling, Uniform Scaling, Diameter Change \\
      Color  & Loose Condition Coloring, Definitive Condition Coloring, Fixed Threshold Coloring \\
      Shape  &  Mark Type Shift, Mark Status Recognition\\
      \revision{Bin} &  Grouping and Color Encoding, Range-Based Grouping, Divide Data with Absolute Attributes, Binary Grouping\\
      \revision{Characterize Distribution} & Overall Data Trend, Comparison \\
      Filter & Definitive Filtering, Elimination, Search Space Adjustment\\
      Select & Tentative Selection, Range-Based Selection, Definitive Selection\\ 
      Navigate & Chart-Body Direction, Chart-Body Distance, Chart Structure\\ 
      Annotate & Place Annotation, Position and Color Adjustment\\ 
      Describe & Brainstorming for Improvement, Illustrating Data Status\\ 
      \revision{Style} & Vague Styling Suggestions, Definitive Styling Instructions\\ 
     \bottomrule
    \end{tabular}
    \caption{Top-N Strategies by Task, where strategies are themes derived from clusters of speech acts sharing similar meanings.}
    
    \label{tbl:Semantic Clusters}
\end{table*} 
}
\section{Results: Analysis Process} \label{chap:analysis}

\subsection{Initial Analysis} \label{chap:initial_analysis}
Before examining the collected utterances in depth, we conducted a brief analysis based on the assumptions in \hyperref[chap:pilot]{Section 3} to better understand the data and refine our analysis process.

\vspace{-5mm}
\begin{figure}[!ht]
  \centering
  \begin{subfigure}[b]{1\columnwidth}
  	\centering
    \includegraphics[width=\textwidth, alt={Big letter A on a gray background.}]{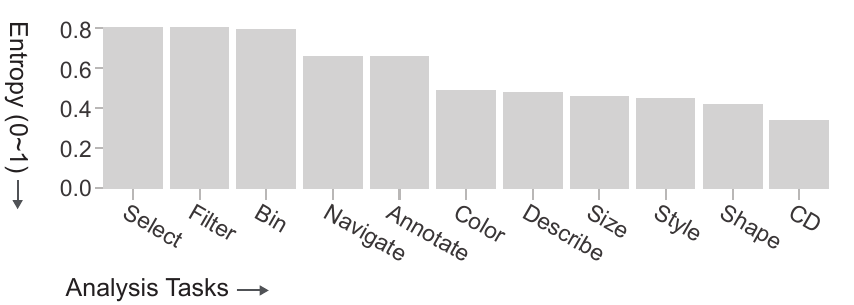}
  	\label{fig:uncertainy_bar}
  \end{subfigure}
\vspace{-7mm} 
  \caption{Speech Input Uncertainty per task, quantified as semantic entropy.}
\end{figure}

\subsubsection{Speech Input Uncertainty}
Following the method of \hyperref[chap:uncertainty]{Section 4.4.2}, we measured \revision{Speech Input Uncertainty} via semantic entropy (\hyperref[fig:uncertainy_bar]{Figure 3}). A significant trend is that data manipulation tasks (\eg Select, Filter, \revision{Bin}, Navigate) mostly exhibit higher entropy. By contrast, encoding and explanation tasks (\eg Color, Size, \revision{Style}, Shape) tend to show lower entropy. We classified \revision{Speech Input Uncertainty} into two levels based on observations: higher entropy ($\geq 0.66$) and lower entropy ($< 0.66$). This suggests a relationship between task type and \revision{uncertainty level}.

\begin{figure}[!htbp]
  \centering
    \begin{subfigure}[b]{1\columnwidth}
  	\includegraphics[width=\textwidth, alt={Big letter A on a gray background.}]{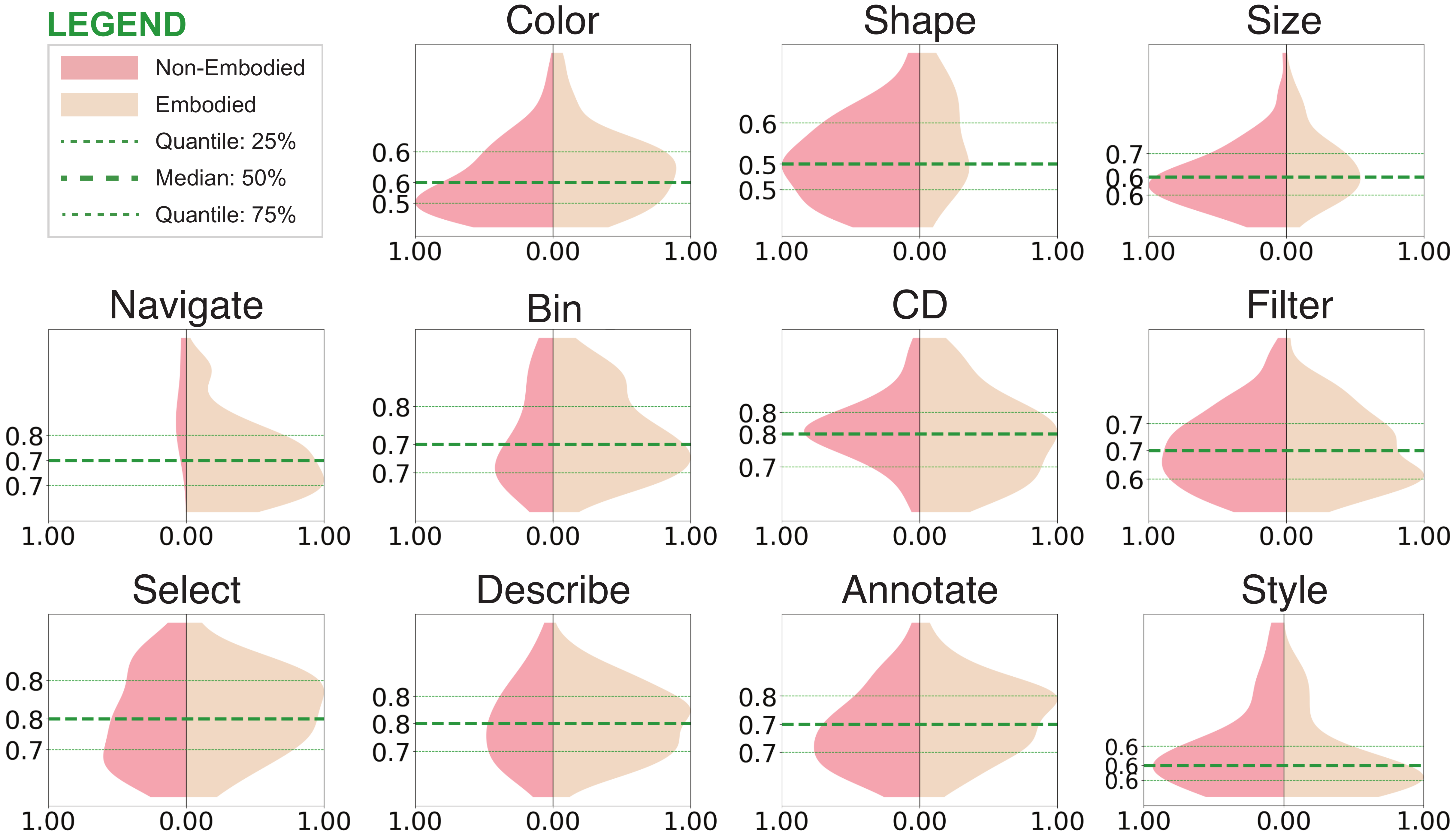}
  	\label{fig:speech_embodied}
    \end{subfigure}
\vspace{-5mm}
  \caption{The plots show the distribution of speech acts based on their semantic embedding distances from the centroid.
1) Y-axis: Semantic distance from the centroid; X-axis: Density of speech acts with the same distance, normalized between 0 and 1.
2) A broader, smoother, right-skewed distribution indicates increased variability and semantic entropy.}
\vspace{-2mm}
\end{figure}

To explore the potential impact of embodiment cues in speech, we categorized speech acts into two groups (\ie Embodied and Non-Embodied) and analyzed their distributions. During axial coding, a speech act tagged with at least one embodiment cue was categorized as an embodied speech act. \hyperref[fig:speech_embodied]{Figure 4} illustrates that embodied speech acts exhibit smoother and broader distributions, reflecting more diverse expressions. In contrast, non-embodied speech acts cluster around peaks that indicate standardized phrasing.

These trends in the charts show that entropy levels and distribution patterns do not always align. For example, tasks such as \textit{\revision{Bin}}, \textit{Filter}, and \textit{Select} exhibit the highest entropy (0.8); however, the distributions of their embodied and non-embodied speech acts differ. In addition, encoding tasks such as \textit{Size, Color, Shape} show lower entropy (0.4-0.6), with embodied speech acts displaying smoother distributions, and non-embodied groups forming sharp peaks. These findings tend to support our initial \hyperref[chap:pilot]{Assumptions (A1, A2)} and point to the need for deeper analysis of how embodied and non-embodied speech acts dynamically interact during task performance.

\vspace{-3mm}
\begin{figure}[!ht]
  \centering
  \begin{subfigure}[b]{1\columnwidth}  
  	\centering
  	\includegraphics[width=\textwidth, alt={Big letter A on a gray background.}]{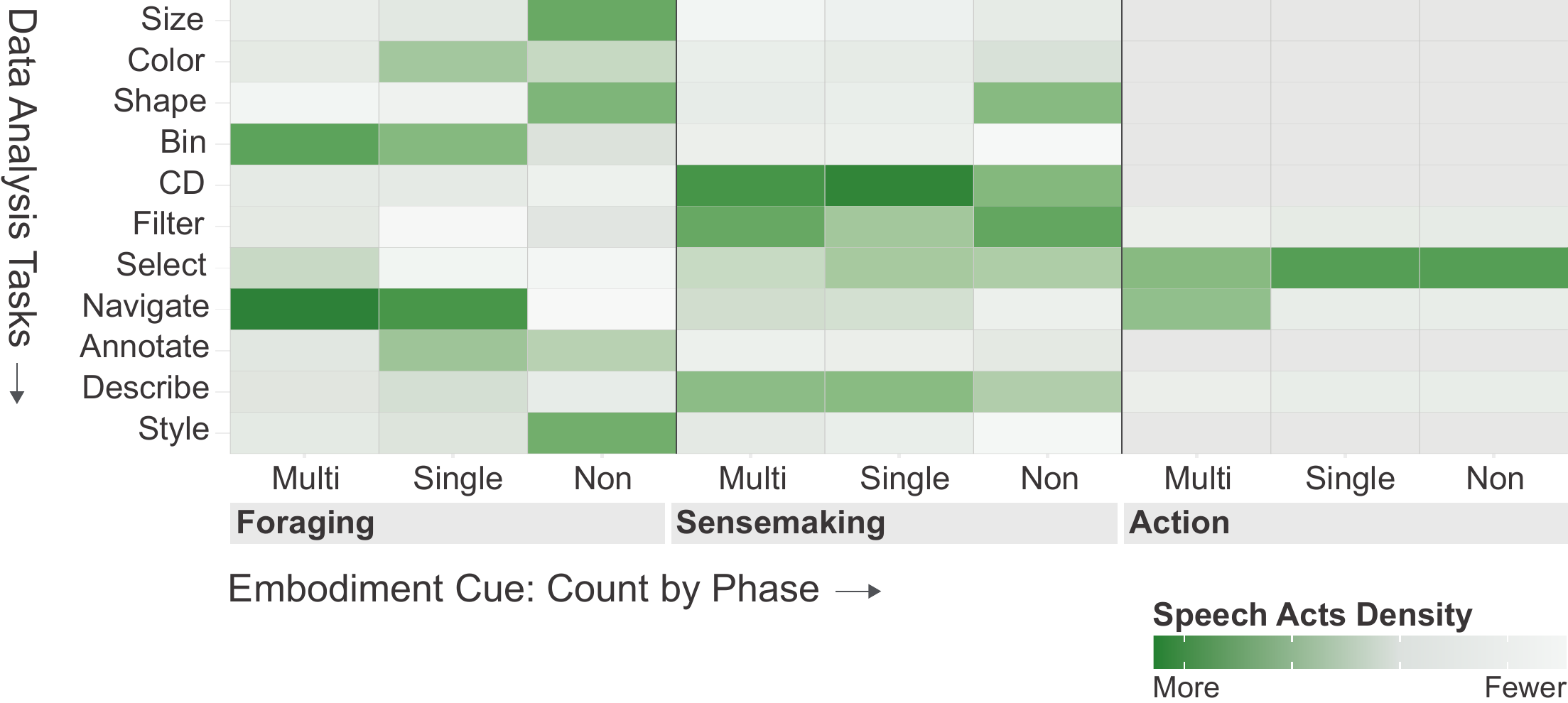}
  	\label{fig:embodiment_cue_count}
  \end{subfigure} 
\vspace{-7mm}
  \caption{Embodiment Cue: Count by Analysis Task and Phase. Each cell indicates the number of speech acts, categorized based on how many embodiment cues are presented within each speech act. Darker colors represent greater reliance on multiple, single, or non-embodiment cues when performing a task.}
\end{figure}

\subsection{Main Analysis} \label{chap:main_analysis}

\subsubsection{Embodiment Reliance}
For the in-depth analysis, we applied three analysis phases and extended the speech act types from two (\ie Embodied and Non-Embodied) to three categories based on embodiment cue count (\ie Multiple, Single, and Non). \hyperref[fig:embodiment_cue_count]{Figure 5} shows the result, highlighting that users may perform identical tasks in a phase using different counts of cues in speech. For example, during \textit{Foraging}, users often perform \textit{\revision{Bin}}, \textit{Navigate}, and \textit{Annotate}, heavily relying on cues. By contrast, they rarely use cues for tasks such as \textit{Size}, \textit{Color}, \textit{Shape}, and \textit{\revision{Style}}, even though these tasks were frequently performed. These trends led us to identify active tasks per phase and compute the \revision{Embodiment Reliance} rate by task and phase. The reliance indicates the proportion of embodied speech acts used for a task in a phase. \hyperref[fig:teaser]{Figure 1} shows the results.



\subsubsection{Five Speech Input Patterns: Speech Input Uncertainty and Embodiment Reliance}
By computing \revision{Embodiment Reliance}, we gauged how strongly users depend on embodiment cues in speech across analysis tasks and phases. This measure serves as a proxy for examining how such cues may contribute to the entropy of utterances by task. We first identified tasks performed at least once within each phase (\eg Action: Select, Navigate, Filter). These tasks were categorized into three sub-groups (i.e., Multi, Single, Non) based on the number of embodiment cues expressed in each speech act. We then retained only the tasks for which at least one sub-group exhibited above-average speech act density, indicating active use. This resulted in a representative set of tasks for each phase. The right side of \hyperref[fig:teaser]{Figure 1} presents these tasks along with their corresponding reliance rates.

Unlike entropy, which clearly separates higher and lower groups (with the lower group ending at 0.49 and the higher group starting at 0.66), the reliance values are more diffusely distributed across three ranges: below 47\%, between 47\% and 55\%, and above 55\%. Based on these ranges, we grouped the tasks across phases.

Interestingly, tasks within each group (e.g., \textbf{G1}) share common characteristics. For example, \textbf{G1}: \textit{Operational} tasks (\ie \revision{Bin}, Navigate, Annotate); \textbf{G2}: \textit{Abstract} tasks (\ie Size, Color, Shape, \revision{Style}); \textbf{G3}: \textit{Reasoning} tasks (\ie Filter, Select); \textbf{G4}: \textit{Data elaboration} tasks (\ie \revision{Characterize Distribution}, Describe, Shape); \textbf{G5}: \textit{Execution} tasks (\ie Select, Navigate). 

In addition, tasks within each group also fall within a consistent \revision{Embodiment Reliance} range (e.g., \textbf{G1} shows higher reliance: Navigate at 97\%, \revision{Bin} at 68\%, Annotate at 55\%). We identified two outliers (i.e., \textbf{G2}: Color at 56\%, \textbf{G4}: Shape at 26\%). We interpret them as the result of participants' passive engagement during the analysis, particularly when covering less frequently used tasks such as encoding. 

Additionally, each task’s entropy scores aligned well with the groupings. For example, three tasks in \textbf{G1} fall in the higher entropy group. These iterative explorations enabled us to understand the possibilities of defining the five patterns considering these two categories: \revision{\textit{Speech Input Uncertainty}} (higher, lower) and \revision{\textit{Embodiment Reliance}} (higher, moderate, lower). Finally, this led to the identification of five distinct patterns: \textbf{Pattern 1}: Higher \revision{Speech Input Uncertainty} and Higher \revision{Embodiment Reliance}, \textbf{Pattern 2}: Lower and Lower, \textbf{Pattern 3}: Higher and Moderate, \textbf{Pattern 4}: Lower and Higher, and \textbf{Pattern 5}: Higher and Lower. \hyperref[fig:teaser]{Figure 1} presents the full context.


\section{Results: Five Patterns and Discussion} \label{chap:results}
    

\subsection{Characteristics of Each Pattern}
\vspace{-3mm}
\begin{figure}[!ht]
  \centering
  \begin{subfigure}[b]{1\columnwidth}
  	\centering
  	\includegraphics[width=8.7cm, alt={pattern1}]{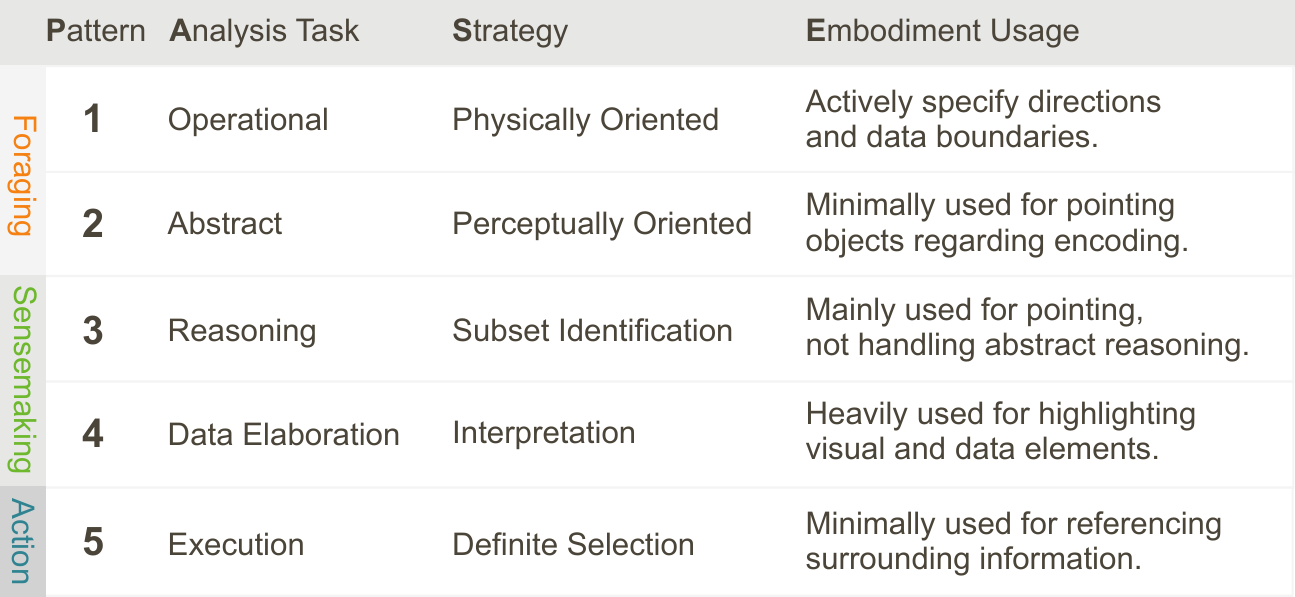}
  	\label{fig:speech_patterns}
  \end{subfigure}
  \caption{Visualization of the five Speech Input Patterns, aligned with corresponding tasks, strategies, and embodiment usage. \hyperref[fig:teaser]{Figure 1} presents the full context of task grouping and pattern distribution.}
\end{figure}

\hyperref[fig:teaser]{Figure 1} and \hyperref[fig:pattern1]{Figure 6} introduce the five \revision{speech input patterns} along with the corresponding tasks, strategies, and users' embodiment usage patterns. In \textit{Foraging}, a sharp contrast appears between \textit{Operational} and \textit{Abstract} tasks, characterized by a positive correlation between \revision{Speech Input Uncertainty (denoted as `Uncertainty' in this section)} and \revision{Embodiment Reliance}. \request{Conversely, in the \textit{Sensemaking} and \textit{Action} phases, the relationship between \revision{these categories} is inconsistent due to several reasons. These dynamics shape the origin of uncertainty differently across phases.}


\subsubsection{Foraging}

\begin{figure}[!ht]
  \centering
  \begin{subfigure}[b]{1\columnwidth}
  	\centering
  	\includegraphics[width=8.7cm, alt={pattern1}]{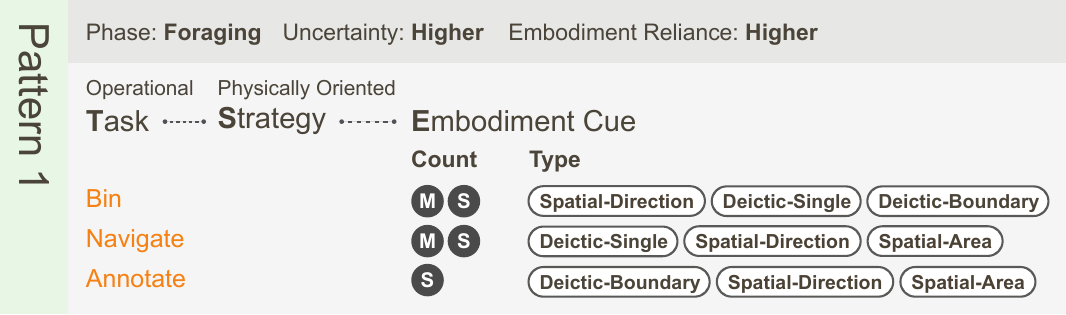}
  	\label{fig:pattern1}
  \end{subfigure}
  \caption{Count indicates the number of embodiment cues appearing in a speech act; \textbf{M}: Multiple, \textbf{S}: Single.}
\end{figure}

\request{
Given \hyperref[fig:teaser]{Figure 1} and \hyperref[fig:pattern1]{Figure 6}, \textit{Operational} tasks—requiring complex spatial manipulation in Foraging—rely more on embodied speech acts (53\% up to over 90\%). In contrast, \textit{Abstract} tasks—focusing on non-spatial properties—rely less on them (up to 52\%). When performing operational tasks, \revision{Speech Input Uncertainty} tends to be higher (entropy $> 0.6$), while it is lower for abstract tasks.
}

\vspace{-3mm}

\bpstart{Pattern 1: Higher Uncertainty and Higher Embodiment Reliance}
\hyperref[fig:pattern1]{Pattern 1} comprises \textit{Operational} tasks performed through \textit{Physically Oriented} strategies (\eg Chart-Body Direction), where spatial adjustments are central. The nature of these tasks actively integrates embodiment cues in speech, primarily to specify objects, directions, and boundaries. \request{In this context, participants adjust spatial properties directly (\eg body and object orientations), which reduces the need for detailed verbal descriptions. Consequently, as participants rely more on embodied speech acts, their utterances tend to become more uncertain. The first example below illustrates that, rather than explicitly verbalizing where they intend to explore, participants integrate hand gestures (\eg pointing above a particular area) into their utterances that contain embodiment cues (e.g., above the mean price). More examples follow (\underline{underlines} indicate embodiment cues).
}
\vspace{-1mm}
\begin{tcolorbox}[highlight]
\begingroup
  \leftskip=0pt
  \noindent\textit{``\underline{Above} the mean price, you paint \underline{those} data points as blue.''} (\revision{Bin})\\
  \noindent\textit{``\underline{It's} a little too \underline{close to me}.''} (Navigate)\\
  \revision{
    \noindent\textit{``Add a new line that highlights houses \underline{which is below}…''} (Annotate)\\
  }%
\endgroup
\vspace{-3mm}
\end{tcolorbox}
\request{This mechanism appears consistently across the five patterns, although with varying degrees of reliance on embodiment. In the Foraging phase, the contrast is particularly pronounced, suggesting that embodied speech acts are the primary factor contributing to elevated \revision{Speech Input Uncertainty}. Further details are provided in \hyperref[chap:pattern1_val]{Section 6.2.1}.}

\begin{figure}[!ht]
  \centering
  \begin{subfigure}[b]{1\columnwidth}
  	\centering
  	\includegraphics[width=8.7cm, alt={pattern1}]{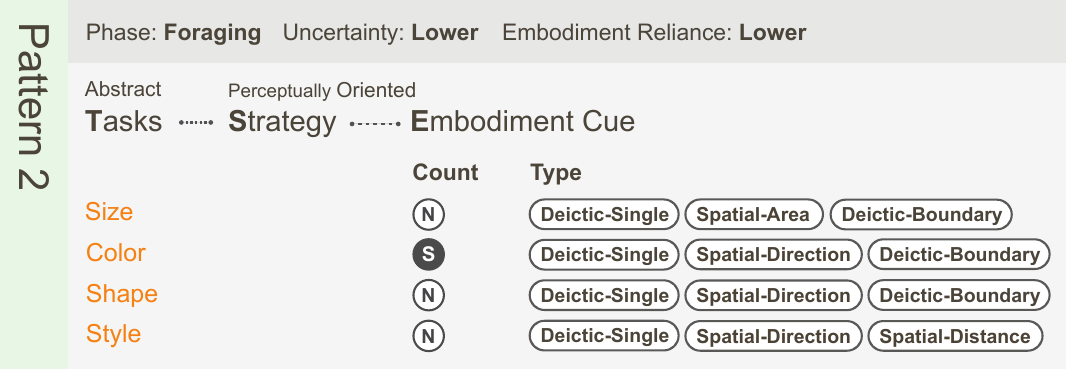}
  	\label{fig:pattern2}
  \end{subfigure}
  \caption{\textbf{S}: Single, \textbf{N}: Non-embodiment cue.}
  \vspace{-3mm}
\end{figure}

\bpstart{Pattern 2: Lower Uncertainty and Lower Embodiment Reliance}
\request{
\hyperref[fig:pattern2]{Pattern 2} illustrates \textit{Abstract} tasks performed using \textit{Perceptually Oriented} strategies (e.g., Conditional Scaling) that show lower variability—for example, subtle condition differences (Size: Conditional Scaling, Uniform Scaling, Diameter Change). These tasks are accomplished through detailed verbal expressions. Such utterances are generally straightforward and standardized, which reduces the need for embodiment cues (\eg pointing to an object).
}
\request{
The examples below illustrate this interplay. The first and second examples use explicit (Size: according to...) or flexible (Color: by using different...) conditions, respectively. Shape and \revision{Style} specify more precise targets for change. This consistency in verbal expressions, combined with minimal spatial referencing, contributes to reducing uncertainty. Therefore, embodiment cues function as supportive references rather than directly shaping the utterance. 
}
\vspace{-2mm}
\begin{tcolorbox}[highlight]
\begingroup
  \leftskip=0pt
  \noindent\textit{``Let's resize \underline{them} according to...''} (Size)\\
  \noindent\textit{``Highlight \underline{them} by using different colors.''} (Color)\\
  \noindent\textit{``Turn the shape of \underline{all} the data points from a cube to a sphere.''} (Shape)\\
  \noindent\textit{``Turn the mean basement marking line from black to green.''} (\revision{Style})
\endgroup
\end{tcolorbox}

\subsubsection{Sensemaking}
Tasks in this phase are performed using both embodied and non-embodied speech acts without strong reliance on either. \textit{Reasoning} tasks (\eg Filter, Select) exhibit higher \revision{Speech Input Uncertainty} and moderate \revision{Embodiment Reliance} (up to 50\%), while \textit{Data Elaboration} tasks (\eg \revision{Characterize Distribution}, Describe) show lower \revision{Speech Input Uncertainty} and higher \revision{Embodiment Reliance} (over 57\%), as shown in \hyperref[fig:teaser]{Figure 1}.

\vspace{-3mm}
\begin{figure}[!ht]
  \centering
  \begin{subfigure}[b]{1\columnwidth}
  	\centering
  	\includegraphics[width=8.7cm, alt={pattern3}]{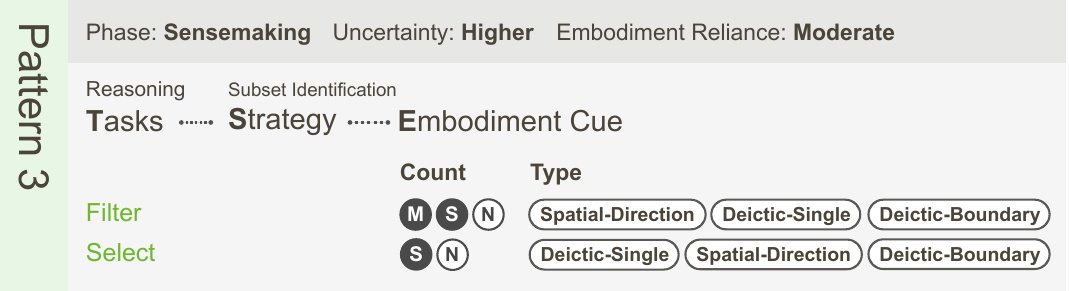}
  	\label{fig:pattern3}
  \end{subfigure}
  \caption{\textbf{M}: Multiple, \textbf{S}: Single, \textbf{N}: Non-embodiment cue.}
\end{figure}

\bpstart{Pattern 3: Higher Uncertainty and Moderate Embodiment Reliance}
\request{\hyperref[fig:pattern3]{Pattern 3} involves \textit{Reasoning} tasks carried out through \textit{Subset-Identification} strategies (\eg Elimination). 
These tasks require sophisticated linguistic expressions, including complex conditions, ambiguity, and nuanced references \cite{baan_uncertainty_2023}, because the tasks entail multiple logical steps during iterative refinements. In this context, embodiment cues mainly anchor references, while linguistic expression serves the logical construction of reasoning to handle abstract concepts. Therefore, as utterances become more sophisticated, they are increasingly peppered with these cues (e.g., after, around) that mark specific points, directions, and data ranges. Specifically, the first example shows that even with several cues, the detailed reasoning process primarily relies on linguistic expressions. As a result, \revision{Embodiment Reliance} remains moderate, while \revision{Speech Input Uncertainty} is higher due to the task characteristics. Further details are in \hyperref[chap:pattern3_val]{Section 6.2.3}.
}

\vspace{-1mm}
\begin{tcolorbox}[highlight]
\begingroup
  \leftskip=0pt
  \noindent\textit{``Eliminate \underline{all} points \underline{after} \$412,000 or \$400,000 and eliminate \underline{all} points \underline{before} the year built 2000.''} (Filter)\\
  \noindent\textit{``I would go with the bigger green ball, or, if not, recommend the first green ball \underline{around} 1,290,420 US dollars.''} (Select)
\endgroup
\end{tcolorbox}

\begin{figure}[!ht]
  \centering
  \begin{subfigure}[b]{1\columnwidth}
  	\centering
  	\includegraphics[width=8.7cm, alt={pattern1}]{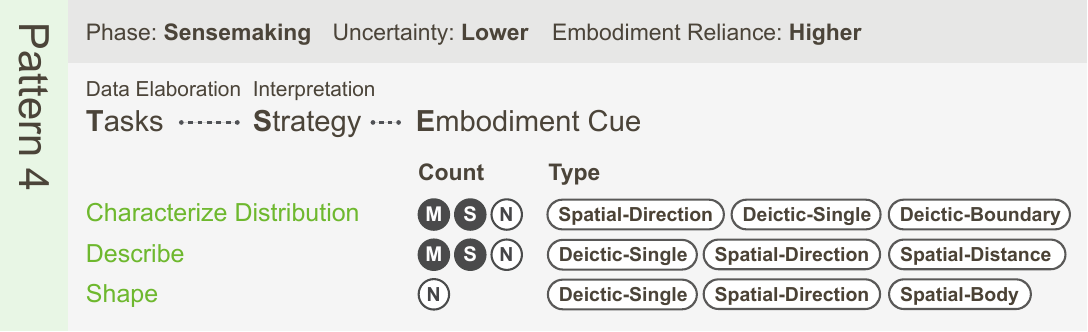}
  	\label{fig:pattern4}
  \end{subfigure}
  \caption{\textbf{M}: Multiple, \textbf{S}: Single, \textbf{N}: Non-embodiment cue.}
\vspace{-3mm}
\end{figure}

\bpstart{Pattern 4: Lower Uncertainty and Higher Embodiment Reliance}
\request{
\hyperref[fig:pattern4]{Pattern 4} involves \textit{Data Elaboration} tasks carried out through Interpretation strategies (e.g., Comparison), in which speech acts primarily request additional information or describe chart observations. In this context, multiple embodiment cues are repeatedly used to explicitly point to specific objects and directions (e.g., these data points, reference line, year built, around 241,000). Another notable feature is that similar sentence structures (e.g., "I see..." "It appears... "It shows...") and comparative terms (e.g., higher than, in the center) are used in speech. These patterns lead to standardized speech, resulting in cohesive and consistent verbal expression.
}

This trend is particularly pronounced in \textit{\revision{Characterize Distribution}} compared to \textit{Describe}. \hyperref[fig:speech_embodied]{Figure 4} shows that the former has a highly centralized distribution of non-embodied speech acts, corresponding to the lowest entropy shown in \hyperref[fig:uncertainy_bar]{Figure 3}. Despite this verbal standardization, the exploratory nature of data elaboration tasks naturally prompted diverse embodied speech acts, reducing cognitive load and enhancing intuitive understanding of the chart. This combination of consistent verbal patterns and strong reliance on embodied speech acts results in lower Speech Input Uncertainty while sustaining high \revision{Embodiment Reliance}. Further details are in \hyperref[chap:pattern4_val]{Section 6.2.4}.


\vspace{-2mm}
\begin{tcolorbox}[highlight]
\begingroup
  \leftskip=0pt
  \noindent\textit{``I can see that \underline{there} are better conditioned houses based on the mean.''} (\revision{Characterize Distribution})\\
  \noindent\textit{``I want to see if \underline{there} are more houses with \underline{lower than}...''} (\revision{Characterize Distribution})\\
  \noindent\textit{``It's clustered \underline{around} one side like it's mostly \underline{between}.''} (Describe)
\endgroup
\end{tcolorbox}
\vspace{-2mm}



\subsubsection{Action}
The Action phase is where users execute the final decision through \textit{Execution} tasks (\eg Select, Navigate) and \textit{Definite Selection} strategies. \request{This phase aims to observe how users execute a final selection rather than interim actions during exploration; thus, we focus on the \textit{Select} task. Given \hyperref[fig:teaser]{Figure 1}, the \textit{Select} task shows lower \revision{Embodiment Reliance} (below 35\%) and higher \revision{Speech Input Uncertainty} (entropy > 0.8).}

\begin{figure}[!ht]
  \vspace{-3mm}
  \centering
  \begin{subfigure}[b]{1\columnwidth}
  	\centering
  	\includegraphics[width=8.7cm, alt={pattern1}]{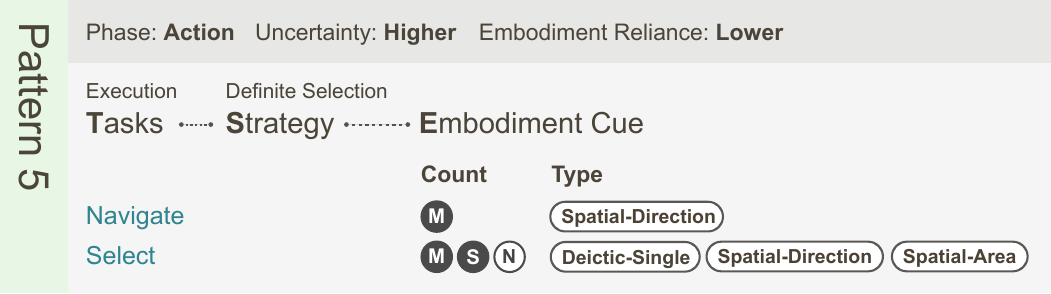}
  	\label{fig:pattern5}
  \end{subfigure}
  \caption{\textbf{M}: Multiple, \textbf{S}: Single, \textbf{N}: Non-embodiment cue.}
\end{figure}

\bpstart{Pattern 5: Higher Uncertainty and Lower Embodiment Reliance}
\request{
In \hyperref[fig:pattern5]{Pattern 5}, embodiment cues appear to reference surrounding information such as an axis, value, or relative location, while linguistic expressions primarily articulate the final target. A common approach is to cite nearby context or annotations (e.g., House IDs displayed next to marks). Notably, some participants even prompted the system to automatically identify the best house. Overall, while participants physically navigated the chart during selection, they primarily relied on non-embodied speech acts to articulate their final choice, resulting in lower \revision{Embodiment Reliance} and higher \revision{Speech Input Uncertainty}. Further details are in \hyperref[chap:pattern5_val]{Section 6.2.5}.
}

\vspace{-1mm}
\begin{tcolorbox}[highlight]
\begingroup
  \leftskip=0pt
  \noindent\textit{``I would choose House ID 67.''} (Select)\\
  \noindent\textit{``The point which is \underline{falling on...}, which I would look at.''} (Select)\\
  \noindent\textit{``Find me a data point that is the most ideal.''} (Select)\\
  \noindent\textit{``The dot \underline{on the right side} would be a bit better.''} (Select)
\endgroup
\end{tcolorbox}


\subsection{Validation through User Reflections}
\request{
This section substantiates the five patterns based on the self-reported responses of participants (denoted as `P').
}
\vspace{-1mm}
\subsubsection{Pattern 1} \label{chap:pattern1_val}
\request{P6 explained their reason for combining embodied speech acts with hand gestures. P8 noted that they resolved navigation challenges by adjusting their body position. P4 indicated they used hand gestures to show where to place annotations.}

\vspace{-1.5mm}
\begin{tcolorbox}[highlight]
\begingroup
  \leftskip=0pt
  \noindent\textit{``I tried to motion what I wanted the graph to do since I couldn’t express it in words.''} (P6, Binning)\\
  \noindent\textit{``I needed to move around to see the chart.''} (P8, Navigate)\\
  \noindent\textit{``I drew the reference lines using gestures while speaking.''} (P4, Annotate)
\endgroup
\vspace{0mm}
\end{tcolorbox}

\vspace{-2mm}
\subsubsection{Pattern 2} \label{chap:pattern2_val}
\request{P6 highlighted that encoding tasks were easy to verbalize, whereas P8’s comment suggested that users might use hand movements to convey details when those tasks involved spatial elements.}

\vspace{-1.5mm}
\begin{tcolorbox}[highlight]
\begingroup
  \leftskip=0pt
  \noindent\textit{“The easiest tasks were to change the colors and the shapes.”} (P6, Shape, Color)\\
  \noindent\textit{“If I want to change the size larger, I'll express it by expanding my hands.”} (P8, Size)\\
\endgroup
\vspace{-3mm}
\end{tcolorbox}

\vspace{-2mm}
\subsubsection{Pattern 3} \label{chap:pattern3_val}
\request{P4 reported confirming specific points by changing their head position while using embodied cues during speech. P12 used hand gestures to indicate data ranges and selection areas when filtering data.}

\vspace{-1.5mm}
\begin{tcolorbox}[highlight]
\begingroup
  \leftskip=0pt
  \noindent\textit{%
    ``I've just moved my head towards the left to see...''} (P4, Select)\\
  \noindent\textit{%
    ``I waved below a certain line to indicate that the system should remove all data points below a certain line.''} (P12, Filter)\\
\endgroup
\vspace{-3mm}
\end{tcolorbox}

\vspace{-2mm}
\subsubsection{Pattern 4} \label{chap:pattern4_val}
\request{P3 used hand movements to confirm the successful execution of commands rather than relying on verbal feedback. P1 also employed pointing gestures when interpreting data trends, prefacing remarks with "I see..." as part of a consistent sentence structure.}

\vspace{-1.5mm}
\begin{tcolorbox}[highlight]
\begingroup
  \leftskip=0pt
  \noindent\textit{%
    ``I was using my hands to explain things. Maybe a thumbs up when my command was completed successfully.''} (P3, Describe)\\
  \noindent\textit{%
    ``I see a bunch of data points that are close together around the five to eleven room count and between \$69{,}882 and \$583{,}726.''} (P1, \revision{Characterize Distribution})\\
\endgroup
\vspace{-3mm}
\end{tcolorbox}

\vspace{-2mm}
\subsubsection{Pattern 5} \label{chap:pattern5_val}
\request{P2 indicated that users naturally lean on embodied actions when selecting an object that appears touchable. P1 pointed out that speech alone is often insufficient for precise selection. Therefore, they incorporated surrounding information as contextual cues in their speech while using embodied actions. These examples show that they rely primarily on linguistic expressions in their final selection.}

\vspace{-1.5mm}
\begin{tcolorbox}[highlight]
\begingroup
\leftskip=0pt
\noindent\textit{%
  ``I tried to touch the dataset I want to see if I can click it and call out the data.  
  Then, I realized I should ask it to label out the ID of the data set.''} (P2, Select)\\
\noindent\textit{%
``When I wanted to short list housing falling within the three variables, I kind of gestured my hand around the lines and said, `Highlight data points below line X'.''} (P1, Select)\\
\endgroup
\vspace{-3mm}
\end{tcolorbox}

\subsection{Challenges and Strengths} \label{chap:RQ2}
Having addressed RQ1 by identifying five patterns that reflect users' approaches, we now turn to RQ2. It is necessary to examine the gap between their expectations and actual behaviors. To explore these gaps, we asked participants about the challenges and strengths they experienced during the study. The corresponding patterns are noted in parentheses after each subject.

\subsubsection{Challenges}
\bpstart{Sophisticated Filtering and Operations (Pattern 1, 3)}
A few participants attempted to apply sophisticated operations and multiple conditions to narrow the exploration space and enhance reasoning. P2 noted a desire to ask the system to perform complex calculations for in-depth analysis, while P10 mentioned difficulties in identifying correlations, which required complex filtering operations and analyses.


\bpstart{System-Understandable vs. Natural Expressions (Pattern 1, 2)}
Several participants reported challenges while verbalizing their intent in ways they thought the system would understand. Although interviewers encouraged them to speak naturally, they often reformulated their utterances concisely by using clear commands and avoiding unnecessary words, sometimes drawing from their previous experience with AI systems. \footnote{(P1, P2, P3, P5, P6, P8, P9, P10)} P1 stated, \textit{``I initially was skeptical about my commands because I did not know if it was accurate enough for the system to take it.''} These tendencies align with the early exploration stages, particularly the \textit{Foraging} phase (Patterns 1 and 2), where users prefer to rely on prior experience and intuition.

\bpstart{Object Selection (Pattern 5)}
Selecting marks was a common challenge. \footnote{(P3, P12)} To explore speech-based selection strategies, we randomly provided two conditions: 1) marks only (early phase) and 2) marks with house IDs (later phase). Participants preferred selecting by ID when available but struggled without IDs, particularly during reasoning and decision-making. In such cases, they referred to surrounding information to describe the target, as discussed in Pattern 5.

\bpstart{3D Chart Literacy (Pattern 1)}
Many participants interacted with the 3D chart based on prior 2D chart experiences. We interpret these observations as issues of 3D chart literacy in immersive environments. Notably, \revision{several} participants did not reposition themselves or move closer to the chart, despite complaining about z-axis readability. \footnote{(P4, P5, P6, P7, P8, P10, P13, P14)} Some mentioned, \textit{``I do not know how to see the z-axis''}, or \textit{``I cannot read text on the z-axis.''} Several kept their distance from the chart, often with arms crossed. Most had little or no XR experience. In contrast, participants with at least basic XR experience intuitively navigated the chart and asked whether they could move around it \footnote{(P1, P11, P14, P15)}. These behaviors highlight how prior XR experience impacts users' ability to \revision{explore} 3D charts during the \textit{Foraging} phase, particularly in relation to Pattern 1.

\bpstart{Undo (Pattern 1, 2, 3, 4)}
Some participants attempted to undo actions by going back to a few steps from the current state. Although not explicitly captured in the five patterns (e.g., Embodiment Cue Type: Spatial-Time), it was observed qualitatively, as participants found it challenging to verbalize such requests (P2, P12). This behavior was especially common during \textit{Foraging} and \textit{Sensemaking}, where participants often wanted to trace changes and return to a previous state while actively exploring.


\bpstart{Feedback (Pattern 1, 2, 3, 4)}
While we focus on speech inputs, \revision{we found that} feedback reduced users’ uncertainty by clarifying command success, data status, and reasons behind visual changes during exploration and decision-making.

\vspace{-1mm}
\subsubsection{Strengths}
\bpstart{Simple Tasks (Pattern 1, 2)}
Participants commonly found simple tasks, such as basic encoding, filtering, and information retrieval, easy to verbalize, with filtering as the most accessible task \footnote{(P4, P6, P7, P10)}. Simple filtering, especially in the early stages, helps them identify interesting data cohorts and increases confidence, as the input-output relationship is straightforward.

\bpstart{Navigation with Body and Speech (Pattern 1)}
A few participants actively navigated both around and within the chart. For example, P14 stated, \textit{``Since it was a 3D visualization, I could actually move around and see the year it was constructed. I could go and see how it extends back as far as 2010. This was helpful for me in filtering out older houses.''} Such feedback from participants directly aligns with Pattern 1, highlighting how body reorientation naturally aided chart navigation and data understanding.

\bpstart{Tutorial (Pattern 1, 2)}
Several participants leveraged the tutorial to learn available commands and adjust their speech, particularly during early exploration in Pattern 1 and 2 \footnote{(P1, P3, P11)}. Some simplified expressions aligned with tutorial examples, suggesting that guidance impacts their expectations of the system's language capabilities.

\revision{However, participants prioritized commands that were aligned with their task needs over strictly following tutorial examples.} For instance, filtering was the most frequently used task during active analysis, even though the tutorial and passive analysis emphasized encoding and styling tasks to collect a balanced range of utterances.

\bpstart{UI and Effect (Pattern 1, 2, 3, 4)}
Participants mentioned that UI panels displaying the live status were helpful. Some noted that animation effects during filtering and sorting helped them better understand the context. These observations align with the \textit{Foraging} and \textit{Sensemaking} phases, where understanding contextual information is essential.


\section{Design Implications}
The design implications leverage the five patterns to discuss designing NLI-based immersive analytic systems focused on speech input under the premise that Generative AI powers the system.

\subsection{General Design Considerations}
\bpstart{Dynamic Uncertainty Management}
Patterns suggest the necessity of dynamically formulating system responses based on analysis phases and tasks. To operationalize this insight in a system, it continuously quantifies Speech Input Uncertainty alongside relevant embodiment cues at each interaction point. These metrics can be used to formulate the system's appropriate response strategy.

\bpstart{Physically Oriented vs. Abstractly Oriented Tasks} 
Immersive visualization tasks fall into two broad categories—physically oriented and abstractly oriented tasks—based on the degree of possible engagement of embodiment, such as the former (e.g., Filter, Navigate). This can serve as a starting point to determine the type of system response to resolve Speech Input Uncertainty. For example, a system resolves ambiguity by providing a visual interface for abstract tasks while encouraging users to use embodiment factors to clarify targets in physically oriented tasks more actively \cite{piumsomboon_grasp-shell_2014}.

\bpstart{Leveraging Spatial Elements}
Immersive analytics enables users to utilize the open-ended space in extended reality compared to traditional visualizations, where information exploration is limited within a screen. As we discussed, the user study also suggested that many users may not know how to leverage these benefits, focusing on speech rather than trying to utilize the spatial elements. As a result, these distinctions should be considered when designing a 3D chart experience in immersive environments.

\subsection{Foraging}
\bpstart{Visual Guidance for Body Reorientation}
The efficiency of combining speech and gestural inputs has been studied in \request{existing traditional and immersive visualization \cite{badam_affordances_2017, batch_wizualization_2024, leon_talk_2024}}. Likewise, guiding users to hidden targets via visual cues improves their data exploration in the foraging phase. Many study participants mentioned difficulties in exploring data along the z-axis, referring to it with casual terms such as \textit{the third line}. When their speech becomes uncertain due to unstandardized terms or deictic expressions while proactively grouping data or displaying a visual object, visual cues indicating the target may help them.

\bpstart{Reactive Visual Annotators}
\request{
Prior research \cite{vaithilingam_dynavis_2024, badam_affordances_2017} has demonstrated the role of affordances. Likewise, in the early stage of interaction, displaying essential affordances near the user's position or in context with the current data analysis can help users learn and formulate speech. Interfaces can be flexibly gathered and arranged  based on speech, aligning with linguistic structures and spatial cues.} For example, once a user starts data exploration with a request, such as \textit{``I am interested in data from here to there,''} indicating two spots with their hands, the system displays visual \revision{anchors} at the pointed locations. These anchors are connected with visual lines and additional options, such as data fields or ranges with aggregated summaries. These cues help users continue to verbalize further pairs of extremes referring to the provided options (\eg \revision{Show} me the average between them.).





\subsection{Sensemaking}
\bpstart{Speech History Visualization} 
\request{
It is well known that tracking user interaction history enables users to perform an \textit{Undo} to refine their analysis strategies \cite{hayatpur_datahop_2020, callahan_vistrails_2006}. Similarly,} embodiment cues and relevant speech acts can be logged with the user’s position and time. This record becomes visible when speech history is enabled, encouraging users to revisit key moments. \request{The logs' visual representation could take the form of} footprint marks indicating position and orientation alongside the chart and data elements used in the reasoning process. For example, users move to a footprint mark when turning on the visualization history. The system then shows the summary of the exploration, such as related chart and speech act information, and asks whether to restart the exploration from the moment. 


\bpstart{Filter Recommendation}
\request{
Complex filtering is a crucial task in data exploration, as examined in prior research \cite{suh_sensecape_2023, song_marrying_2024}. To support efficient filter composition in a speech-driven immersive analytics system, the system should adaptively suggest additional filters in conjunction with the user's spoken inputs and expected outcomes.} For example, when a user asks \textit{``What is the most recently built house?''}, the system highlights the house on the chart. Two connected physical nodes appear beside the user, presenting recommended filters: Node 1 (\textit{most recently}) means an operation, while Node 2 (\textit{built in 2022}) indicates a specific data field. Additional node pairs appear below, such as: Node 1: \textit{Oldest}, Node 2: \textit{Built in 2001}, Node 3: \textit{Room Count is under 4}. Through this process, users gradually learn how to formulate verbal expressions and available data constraints across the foraging and sensemaking phases.


\bpstart{Uncertain Expression Identifier}
\request{
Disambiguating user intent has been explored in visualization research through the use of predefined logic and interfaces \cite{gao_datatone_2015, setlur_eviza_2016}. Leveraging AI capabilities, the approach can operate on the fly with speech input.
An utterance naturally involves embodiment cues. Thus, chunks of the utterance—including the cues—should be actively used to guide users as they brainstorm the next utterance or gesture in real time. For instance, the system displays the utterance (\textit{\eg the trend \underline{near the x-axis} is similar to \underline{here}}) in bottom, highlighting words or phrases that vaguely articulate data and chart elements (\textit{\eg the trend \underline{near the x-axis}}). The system provides textual details and confidence scores for each highlighted one, offering transparency about potential ambiguity. In addition, the chart visually highlights the targets, showing probable intervals. Referring to the details, users formulate the next utterance or re-point to the target.}



\subsection{Action}
\bpstart{Selection Facilitator} The system switches to the in-depth selection mode to facilitate definitive selection, removing less relevant data points from the chart and showing object indicators near them. Simultaneously, it visually emphasizes objects based on the proximity to the user's position. Referring to both factors, the user verbalizes the final choice.



\section{Limitations and Future work}
\revision{
One limitation is the imbalanced sample sizes per task. We aimed for a balanced distribution of utterances across analysis tasks through passive analysis; however, users' task preferences \revision{(\eg Filter)} resulted in an imbalance. Addressing this requires additional data collection or careful resampling across tasks.}
\request{For a future study, we could add more chart types to broaden the use cases. However, as noted in \hyperref[chap:scatterplot]{Section 4.3}, we intentionally focused only on scatterplots. Therefore, extending the chart library to include additional chart types—which would require substantial development effort—is beyond the scope of this research.}
\revision{Additionally, future studies could incorporate gesture data.} In our study, we did not incorporate gestures; instead, we focused on embodiment cues embedded in utterances. \revision{Including gestural data is a possible next step. We experimented} by adding participants' gestures captured from video recordings to the current coding results. This data augmentation may serve as a foundation for designing multimodal interactions for immersive visualizations.
\revision{Lastly, we are interested in designing a speech-driven immersive analytics system that leverages generative AI, applying the five patterns and associated insights from this study.}

\section{conclusion}
This study identifies five speech input patterns\revision{—shaped by Speech Input Uncertainty and Embodiment Reliance—that users dynamically blend in immersive analytics. The results inform design implications. Future work will examine the role of natural language in generative AI-infused speech-driven immersive analytics systems.}

\section*{Supplemental Materials}
\label{sec:supplemental_materials}
All supplemental materials are available at our OSF repository \href{https://osf.io/sv4fn/}{(https://osf.io/sv4fn/)}. They include:  
(1) Axial coding results  
(2) Exit-interview data  
(3) Analysis by task (example task: color)  
(4) Photos (wizard workstation, installation, user-study setup)  
(5) Pilot-study documentation.  






\acknowledgments{%
We deeply appreciate the constructive comments from the anonymous reviewers of this manuscript, which helped improve the quality of this work. This research was funded by the US Government contract H98230-19-D-0003/0027. Any opinions, findings, conclusions, or recommendations expressed in this article are those of the authors and do not necessarily reflect the views of the research sponsors, nor does mention of trade names, commercial products, or organizations imply endorsement by the US Government.%
}

\bibliographystyle{abbrv-doi-hyperref}

\bibliography{embodiednli}








\end{document}